\def\fr#1#2{\hbox{${#1\over #2}$}}

\def\+{{(+)}}  \def\-{ {(-)} }   \def\0{ {(0)} } 
\def\1{ {(1)} }  \def\2{ {(2)} }

                            \def\k{\kappa}

\documentclass[11pt]{article}

\def\cleq{\setcounter{equation}{0}} 
 
\def\be{\begin{equation}}             \def\ee{\end{equation}} 
\def\ba{\begin{array}{rcl}}           \def\ea{\end{array}} 
\def\beqa{\begin{eqnarray} }          \def\eeqa{\end{eqnarray} } 
\def\beqalign{\begin{eqalign}}        \def\eeqalign{\end{eqalign}} 
             \def\eq#1{(\ref{eq:#1})} 
\def\bsubeq{\begin{subequations}}     \def\esubeq{\end{subequations}} 
\def\bitem{\begin{itemize}}           \def\eitem{\end{itemize}}

\def\DJ{\leavevmode\setbox0=\hbox{D}\kern0pt 
 \rlap{\kern.04em\raise.188\ht0\hbox{-}}D} 
\def\dj{\leavevmode\setbox0=\hbox{d}\kern0pt 
 \rlap{\kern.215em\raise.46\ht0\hbox{-}}d}

\newcommand{\bd}{\begin{displaymath}} 
\newcommand{\ed}{\end{displaymath}} 
 
\begin{document} 
\title{Thirring sine-Gordon relationship by canonical 
methods} 
\author{V.Juri\v ci\' c$^*$ and B. Sazdovi\' c$^\dagger$ \medskip \\ 
{\small \textit{$^*$ Department of Physics, University of Fribourg, P\'erolles, 
CH-1700 Fribourg, Switzerland}}\\ {\small \textit{$^*$ Institute of Physics, 
P.O.Box 57, 11001 Belgrade, Yugoslavia}}\\  {\small 
\textit{$^\dagger$Institute of Physics, P.O.Box 57, 11001 Belgrade, 
Yugoslavia}}} 
 
\date{} 
 
\maketitle 
 
\begin{abstract} 
 Using the canonical 
method developed for anomalous theories, we present the independent 
rederivation 
 of the quantum relationship between the massive Thirring and the 
sine-Gordon models. The same 
 method offers the possibility to obtain the 
Mandelstam soliton operators as a solution of 
 Poisson brackets "equation" for 
the fermionic fields. We checked the anticommutation 
 and basic Poisson 
brackets relations for these composite operators. The transition from the 
Hamiltonian to the corresponding Lagrangian variables produces the known 
Mandelstam's result. 
\end{abstract} 
 
\section{Introduction} 
   The connection between massive Thirring model of interacting fermions 
and  sine-Gordon model with 
 nonlinear scalar field is well known 
\cite{Kolman,Mandelstam,stone}. This Bose-Fermi equivalence has been 
obtained   in  \cite{Kolman} by performing the computations of the Green's 
functions for both  theories. After identification of   some parameters, the 
Green's functions  became equal perturbation series, so that under these 
 conditions this two 
theories are identical. An important step towards the obtainig of this result 
has been achieved in a pioneering paper \cite{wilde}. In Ref. 
\cite{Mandelstam} Mandelstam has constructed  the fermi fields as nonlocal 
functions of the sine-Gordon scalars. He showed  that   the corresponding 
operators create and annihilate the bare sine-Gordon  solitons. These 
operators satisfy the   proper commutation relations as well 
as the Thirring model field equations, which confirms Coleman's result. This 
equivalence has been established on the quantum level and the relation between 
the Fermi and Bose fields is non-local. Beside the approaches mentioned 
above,  Fermi-Bose equivalence was obtained in Refs. \cite{halpern} and 
\cite{Morchio} using quantum mechanical interaction picture and 
Krein realization of the massless scalar field. The same problem has been considered 
in the papers \cite{stoyanov}. 
 
 In this paper we 
are going to  derive above connection between massive Thirring and sine-Gordon models using 
canonical method \cite{HT,MS,Sazdovic}. Starting with the fermionic Thirring  model we are going to 
construct the equivalent bosonic theory, which appears  to be the sine-Gordon one. Our approach is 
different from the   previously mentioned ones and naturally works in the {\it Hamiltonian} 
formalism.   It gives a simpler proof of the same result. 
 
We consider the formulation of the Thirring model with auxiliary vector fields,
which on the equations  of motion give the standard form of the Thirring model
action. It is more convenient, regarding the  fact that the method, which we
use, is based on canonical formalism. The form of the action with  auxiliary
fields becomes linear in the fermionic current $j_\mu$.   
 In Sec. 2.1 we are going to canonicaly quantize the 
fermionic field. So, it is useful 
 to keep all parts which contain this field 
and to omit the bilinear part in auxiliary field. Such 
 Lagrangian is 
invariant under local abelian gauge transformations. Consequently, the first 
class 
 constraints (FCC) $j_{\pm}$ are present in the theory and satisfy 
abelian algebra as a Poisson bracket (PB) algebra. 
 The quantum theory is 
anomalous, so that the {\it central term} appears in the commutator algebra of 
the 
 operators ${\hat j}_{\pm}$, and the constraints become second class 
(SCC). 
 
 We define the effective bosonized theory, as a classical theory 
whose PB algebra of the constraints $J_{\pm}$ is isomorphic to the 
 commutator 
algebra of the operators ${\hat j}_{\pm}$, in the 
 quantized fermionic theory. 
Also, the bosonic Hamiltonian depends on $J_{\pm}$ in the same way as the 
fermionic Hamiltonian depends on $j_{\pm}$. The bosonized theory 
 incorporates 
anomalies of the quantum fermionic theory at the classical level. 
 
 In Sec. 2.3 we find the effective Lagrangian for given algebra as its 
 PB current 
algebra and given Hamiltonian in terms of the currents. The similar problems 
have been solved before in the literature \cite{Sazdovic} using canonical 
methods. 
 We introduce the phase space coordinates $\varphi , \pi$ and 
parameterize the constraints $J_{\pm}$ by them. Then we find the expressions 
for the 
 constraints $J_{\pm}$ in terms of phase space coordinates, satisfying 
given 
 PB algebra, as well as for the Hamiltonian density ${\cal H}_c$. We 
then use the general canonical 
 method \cite{HT,MS,Sazdovic} for constructing 
the effective Lagrangian 
 with the known representation of the constraints. Eliminating the momentum 
 variable on its equation of motion we obtain the 
bose theory 
 which is equivalent to the quantum fermi theory. Finally, 
returning omitted term, bilinear in 
 $A_\mu$ and eliminating auxiliary 
vector field on its equation of motion we obtain the sine-Gordon model. 
 
 By 
the way, we obtain Hamiltonian bosonization formulae for the currents which 
depend on 
 the momenta, while those for scalar densities depend only on the 
coordinates. Known Lagrangian  bosonization rules can be 
 obtained from the 
Hamiltonian ones, after eliminating the momenta.
 
The massless Thirring model is considered separately in Sec. 2.4. It is shown
that in its quantum action exists one parameter which does not appear in the
classical one. Therefore, the quantum massless Thirring model is non-uniquely
defined in agreement with Ref. \cite{Hagen}.     

In Sec. 3. the same method 
will be applied for the construction of the fermionic Mandelstam's  
operators.  The algebra of the currents is the basic PB algebra. Commutation
relations  between the currents 
 and fermionic fields completely define the fermionic 
fields. So, we first find the PB between 
 $j_\pm$ and $\psi_\pm$. The 
corresponding commutation relations of the operators ${\hat j}_{\pm}$ and 
${\hat \psi}_\pm$ are not anomalous. In order to obtain the bosonized 
expression for the  fermions 
 we "solved" PB equation, which is isomorphic to the previous 
operator relation. 
 We find the representation for the unknown fermion field 
using the known representation for the currents $J_{\pm}$. 
 The solution 
depends on the phase space coordinates $\varphi$ and $\pi$ and represents 
Hamiltonian form of the 
 Mandelstam's creation and annihilation operators. 
 
Sec. 4. is devoted to concluding remarks. The derivation of the 
 central term, 
using normal ordering prescription, is presented 
 in the Appendix A, and the 
field product regularization in the Appendix B. 
 
\section{Thirring Model} 
 
In this section the canonical method of bosonization will be applied to
the Thirring model.  
 
\subsection{Canonical analysis of the theory} 
 
Thirring model \cite{Tiring} is a theory of massive Dirac field in two--dimensional space--time defined by the 
following Lagrangian 
\be \label {eq:thi} 
\mathcal{L}_{Th} = {\bar \psi} (i \gamma^{\mu} \partial_{\mu} - m ) \psi -\frac{g}{2} j_\mu j^\mu, 
\ee 
where $g$ is coupling constant, and $j^\mu\equiv\bar{\psi}\gamma^\mu\psi$ is the fermionic current. 
In two--dimensional space--time $\gamma$ -- matrices are 
defined in terms of Pauli matrices $\sigma_1$, $\sigma_2$ and $\sigma_3$ as 
$\gamma^0=\sigma_1,\,\, \gamma^1=-i\sigma_2,\,\, 
\gamma_5=-i\sigma_1\sigma_2=\sigma_3 $ and obey standard relations 
\bd 
\gamma^\mu \gamma^\nu + \gamma^\nu \gamma^\mu = 2\eta^{\mu\nu} \,   , \qquad 
\gamma^{\mu}\gamma_5 + \gamma_5\gamma^{\mu}=0   \,  . 
\ed 
Metric tensor $\eta^{\mu\nu}$ is defined by  $\eta^{00}=-\eta^{11}=1$; 
$\eta^{01}=\eta^{10}=0$. Axial--vector product $\gamma^{\mu}\gamma_5$ can be 
expressed in terms of $\gamma^{\nu}$ in a following way 
\bd 
\gamma^\mu \gamma_5 = -\epsilon^{\mu\nu}\gamma_\nu, 
\ed 
where $\epsilon^{\mu\nu}$ is totally antisymmetric tensor $\epsilon^{01}=-\epsilon^{10}=1$. 
Weyl or chiral spinors are defined using $\gamma_5$ matrix 
\bd 
\gamma_5 \psi_{\pm}=\mp\psi_{\pm}, 
\ed 
which can be expressed with the help of chiral projectors 
$P_{\pm}\equiv   \frac{1\mp\gamma_5}{2}$ as 
\bd 
P_{\pm}\psi_{\pm}= \pm\psi_{\pm}. 
\ed 
The definition of the projectors  $P_{\pm}$ implies that 
Dirac spinor $\psi$ expressed in terms of Weyl spinors $\psi_{\pm}$ 
has a form 
\bd 
 \psi=\left (\begin{array}{c} 
\ \psi_{-} \\ 
 \psi_{+} 
 \end{array}\right). 
\ed 
Lagrangian given by Eq. (\ref {eq:thi}) is on--shell equivalent to the following one 
\be \label{eq:vthi} 
\mathcal{L} = \bar{\psi}(i\gamma^{\mu}\partial_{\mu} -  m)\psi + \frac{1}{2} j^\mu A_\mu + 
\frac{1}{8g} A^\mu A_\mu. 
\ee 
Namely, equations of motion for the auxiliary field $A_{\mu}$, which are obtained from 
Lagrangian (\ref {eq:vthi}), have the form 
\bd 
\frac{1}{2}j^\mu + \frac{1}{4g} A^\mu = 0, 
\ed 
which, after substitution in  (\ref {eq:vthi}), gives Lagrangian (\ref {eq:thi}). 
 
We are going to quantize the fermionic field, so we will consider the Lagrangian 
\be \label{eq:vthi0} 
\mathcal{L}_0 = \bar{\psi}(i\gamma^{\mu}\partial_{\mu} -  m)\psi + \frac{1}{2} j^\mu A_\mu  \,  , 
\ee 
keeping the terms with fermionic fields. The canonical method of bosonization will be applied to this 
Lagrangian. In terms of Weyl spinors $\psi_{\pm}$ and light--cone components of auxiliary field $A_\mu$ it reads 
\bd 
\mathcal{L}_0= 
i\psi^*_{-}\dot{\psi}_{-}+  i\psi^*_{+}\dot{\psi}_{+}+ 
i\psi^*_{-}{{\psi}'_{-}}- i\psi^*_{+}{{\psi}'_{+}} - m (\psi^*_{-}\psi_{+} + 
\psi^*_{+}\psi_{-}) 
\ed 
\be 
\label {eq:kthi} + 
\frac{1}{2}(j_{+}A_{-}+j_{-}A_{+}) \,    , 
\ee 
where the chiral currents $j_{\pm}$ are defined by 
$j_{\pm}\equiv\sqrt{2}\psi^*_{\pm}\psi_{\pm}$, and fields
$A_{\pm} \equiv (1/\sqrt{2})(A_0 \pm A_1)$.  Time and space coordinate, are
respectively $\tau \equiv x^0$ and $\sigma \equiv x^1$, and corresponding
derivatives are  $\dot{\psi}\equiv\frac{\partial\psi}{\partial\tau}$  and 
$\psi'\equiv\frac{\partial\psi}{\partial\sigma}$.   
Now we will investigate Hamiltonian structure of the theory defined 
by Lagrangian (\ref {eq:kthi}). This Lagrangian is already in 
Hamiltonian  form. It is linear in time derivatives of basic Lagrangian variables $\psi_{+}$ and $\psi_{-}$, 
whose conjugate momenta are $\pi_{\pm}=i\psi^*_{\pm}$. Variables 
without time derivatives, $A_{+}$ and $A_{-}$, are Lagrange multipliers and 
the primary constraints corresponding to them are the FCC 
\be \label {eq:j} 
j_{\pm}\equiv \sqrt{2}\psi^*_{\pm}\psi_{\pm}=-i 
\sqrt{2}\pi_{\pm}\psi_{\pm}. 
\ee 
 
From Eq. (\ref {eq:kthi}) we can conclude that canonical Hamiltonian density of the Thirring model takes a form 
\be 
\mathcal{H}_c = -i(\psi^*_{-}{{\psi}'_{-}}- \psi^*_{+}{{\psi}'_{+}})+ m 
(\psi^*_{-}\psi_{+} + \psi^*_{+}\psi_{-}) = t_+ - t_- +  m (\rho_{+} + \rho_{-}) \,    , 
\ee 
where we introduced the energy--momentum tensor $t_{\pm}$ and chiral densities 
$\rho_{\pm}$ by the relations 
\bd 
t_{\pm}\equiv 
i\psi^*_{\pm}\psi'_{\pm}=\pi_{\pm}\psi'_{\pm} \,  ,  \qquad 
\rho_{\pm}\equiv \psi^*_{\pm}\psi_{\mp}= -i \pi_{\pm}\psi_{\mp}  \,   . 
\ed 
Total Hamiltonian is defined as 
\be  \nonumber H_T = \int d\sigma 
\mathcal{H}_T, 
\ee 
where total Hamiltonian density, $\mathcal{H}_T$, is 
\be \label  {eq:tht} 
\mathcal{H}_T = t_{+}-t_{-} + m (\rho_{+}+\rho_{-})  \label {eq:tht} - \frac{1}{2} 
(j_{+}A_{-}+j_{-}A_{+})   \,  . 
\ee 
Starting with basic PB 
\be 
\label {eq:tspz} 
\{\psi_{\pm}(\sigma),\pi_{\pm}(\bar{\sigma})\}=\delta(\sigma-\bar{\sigma}) \,  , 
\ee 
it is easy to show that currents $j_{\pm}$ satisfy two independent abelian PB 
algebras 
 \be \label {eq:taj}  \{j_{\pm}(\sigma), 
j_{\pm}(\bar{\sigma})\}= 0\;\;,\;\;\{j_{+}(\sigma),  j_{-}(\bar{\sigma})\}=0. 
\ee 
Using Eq. (\ref {eq:tspz}), we can find PB of the currents $j_{\pm}$ with 
the quantities $t_{\pm}$ and $\rho_{\pm}$ 
\be \{j_{\pm}(\sigma), t_{\pm}(\bar{\sigma})\}= 
-j_{\pm}(\sigma) \delta'(\sigma-\bar{\sigma})\;\;,\;\;\{j_{\pm}(\sigma), 
t_{\mp}(\bar{\sigma})\}=0, 
\ee 
\be \{j_{\pm}(\sigma), 
\rho_{\pm}(\bar{\sigma})\}=-i\sqrt{2}\rho_{\pm}\delta(\sigma-\bar{\sigma}),\hspace{.7cm} 
\{j_{\pm}(\sigma), 
\rho_{\mp}(\bar{\sigma})\}=i\sqrt{2}\rho_{\mp}\delta(\sigma-\bar{\sigma}). 
\ee 
The last relations imply 
\be \label {eq:thcj} 
\{\mathcal{H}_c(\sigma),j_{\pm}(\bar{\sigma})\}=\pm 
j_{\pm}(\sigma)\delta'(\sigma-\bar{\sigma})\mp i m 
\sqrt{2}(\rho_{+}-\rho_{-})   \,  , 
\ee 
which help us to obtain 
\be \label {eq:tvj} 
\dot{j}_{+}=\{j_{+},H_T\}= j'_{+}+i m 
\sqrt{2}(\rho_{+}-\rho{-}),\ee \be \label {eq:tvjj} 
\dot{j}_{-}=\{j_{-},H_T\}= - j'_{-}-i m 
\sqrt{2}(\rho_{+}-\rho{-}).\ee 
Taking the sum of the equations (\ref {eq:tvj}) and (\ref {eq:tvjj}), we get 
\be 
\partial_{-}j_{+} + \partial_{+}j_{-}= \partial_\mu j^\mu =0, 
\ee 
which implies that the current $j^{\mu}$ is conserved. 
 
Since the constraints $j_{\pm}$ are the first class ones, classical theory, defined by Lagrangian \eq{vthi0}, 
has local abelian symmetry, whose generators $j_{\pm}$ satisfy abelian PB 
algebra, given by Eq. (\ref {eq:taj}). 
 
\subsection{Quantization of the fermionic theory} 
 
Passing from the classical to the quantum theory can be obtained by introducing 
the operators $\hat{\psi}$ and $\hat{\pi}$, instead of corresponding classical 
fields. Poisson brackets are replaced by corresponding commutators, and 
operator product is defined using normal ordering prescription, whose details 
are explained in Appendix A, 
\bd \hat{j}_{\pm}\equiv 
-i\sqrt{2}:\hat{\pi}_{\pm}\hat{\psi}_{\pm}:,\;\;\hat{t}_{\pm}\equiv\, 
:\hat{\pi}_{\pm}\hat{\psi}'_{\pm}:,\;\;\hat{\rho}\equiv\,-i :\hat{\pi}_{\pm} 
\hat{\psi}_{\mp}: \,\,.\ed 
Algebra of the operators $\hat{j}_{\pm}$, $\hat{t}_{\pm}$ i $\hat{\rho}_{\pm}$ 
takes a form (Appendix A) 
\be 
\label {eq:qj}\ 
[\hat{j}_{\pm}(\sigma),\hat{j}_{\pm}(\bar{\sigma})]= \pm 
2i\hbar\kappa\,\delta'(\sigma-\bar{\sigma}), \hspace{1cm} 
[\hat{j}_+ (\sigma),\hat{j}_{-}(\bar{\sigma})]=0 
\ee 
\be 
\label {eq:qt} \ 
[\hat{j}_{\pm}(\sigma),\hat{t}_{\pm}(\bar{\sigma})]= -i\hbar 
\hat{j}_{\pm}(\bar{\sigma})\delta'(\sigma-\bar{\sigma}), 
\hspace*{1cm} [\hat{j}_{\pm}(\sigma),\hat{t}_{\mp}(\bar{\sigma})]=0, 
\ee 
\be 
\label {eq:qro} \ [\hat{j}_{\pm}(\sigma),\hat{\rho}_{\pm}(\bar{\sigma})]= 
\hbar \sqrt{2}\hat{\rho}_{\pm}\delta(\sigma-\bar{\sigma}), 
\hspace*{.5cm} 
[\hat{j}_{\pm}(\sigma),\hat{\rho}_{\mp}(\bar{\sigma})]= -\hbar 
\sqrt{2}\hat{\rho}_{\mp}\delta(\sigma-\bar{\sigma}), 
\ee 
where $\kappa\equiv\frac{\hbar}{2\pi}$. 
 
The current operators with different chirality commute, as well as the 
corresponding variables in the classical theory. The difference between classical 
and quantum  algebra is appearance of the central term in the commutator current algebra (\ref {eq:qj}). As 
its consequence, the operators $\hat{j}_{\pm}$ are the second class constraints 
operators. This leads to the existence of the anomaly. Namely, symmetry of the 
classical theory, whose generators are the first class constraints $j_{\pm}$, 
is no longer symmetry at the quantum level. 
 
\subsection{Effective bosonic theory} 
 
Now we will introduce new variables $J_{\pm}, \Theta_{\pm}, 
R_{\pm}$ and postulate their PB algebra to be isomorphic to commutator algebra 
in the quantum fermionic theory, given by Eqs. (\ref {eq:qj}), (\ref 
{eq:qt}), and (\ref {eq:qro}) 
\be \label {eq:alj} 
 \ \{J_{\pm}(\sigma),J_{\pm}(\bar{\sigma})\}= 
\pm 2\kappa\delta'(\sigma-\bar{\sigma}),\hspace{.6cm} 
\{J_{+}(\sigma),J_{-}(\bar{\sigma})\}=0, 
\ee 
\be \label {eq:alt} \ 
\{\Theta_{\pm}(\sigma),J_{\pm}(\bar{\sigma})\}= 
J_{\pm}(\sigma)\,\delta'(\sigma-\bar{\sigma}), \hspace{.5cm} 
\{\Theta_{\pm}(\sigma),J_{\mp}(\bar{\sigma})\}=0, 
\ee 
\be \label {eq:alr} 
\{J_{\pm}(\sigma), R_{\pm}(\bar{\sigma})\}= - i \sqrt{2}R_{\pm}(\sigma)\delta(\sigma-\bar{\sigma}), \hspace{.2cm} 
\{J_{\pm}(\sigma), R_{\mp}(\bar{\sigma})\}= i \sqrt{2}R_{\mp}(\sigma)\delta(\sigma-\bar{\sigma})  \,  . 
\ee 
 
Let us find the expressions for the currents $J_{\pm}$, energy--momentum tensor 
$\Theta_{\pm}$, and chiral densities $R_{\pm}$ in terms of scalar field 
$\varphi$ and its conjugate momenta $\pi$, with the PB 
\bd 
\{\varphi(\sigma),\pi(\bar{\sigma})\}=\delta(\sigma-\bar{\sigma}). 
\ed 
Assuming that the currents $J_{\pm}$ are  linear in the momentum $\pi$, it is easy to 
show that the expression 
\begin {equation} \label {eq:sjpm} 
\ J_{\pm}= \pm \pi + \kappa \varphi' 
\end{equation} 
is a solution of (\ref {eq:alj}). Supposing that the energy--momentum tensor $\Theta_{\pm}$ is quadratic in 
the  currents $J_{\pm}$, we can immediately obtain its bosonic representation 
from algebra (\ref {eq:alt}) 
\begin {equation} \label {eq:steta} 
 \Theta_{\pm}=\pm \frac{1}{4\kappa} J_{\pm} J_{\pm}. 
\end{equation} 
 
Bosonic representation for scalar densities, $R_{\pm}$, can be 
obtained from algebra (\ref {eq:alr}). Assuming that scalar densities are 
momentum independent, we have 
\be 
\label {eq:ro} R_{\pm}=M \exp{(\pm i \sqrt{2}\varphi)}, 
\ee 
where M is constant. The scalar densities as well as parameter M have dimension of mass. 
 
Total Hamiltonian density of the effective bosonic theory is defined by 
the analogy with total Hamiltonian density of the fermionic theory, 
given by Eq. (\ref {eq:tht}) 
\be 
\label {eq:ham} \mathcal{H}_T = \Theta_{+}-\Theta_{-}+ m (R_{+}+ 
R_{-})-\frac{1}{2}(J_{+}A_{-}+J_{-}A_{+})   \,  , 
\ee 
and the Lagrangian of the effective bosonic theory has the form 
\be \label {eq:lagr} 
\mathcal{L}^{Th}_{0}=\pi\dot{\varphi}-\mathcal{H}_T  \,   . 
\ee 
Substituting Eq. (\ref {eq:ham}) in Eq. (\ref {eq:lagr}), and using 
Eqs. (\ref{eq:sjpm}), (\ref {eq:steta}) and  (\ref {eq:ro}), we obtain 
\be \label {eq:elagr} 
\mathcal{L}^{Th}_{0}=\pi\dot{\varphi}-\frac{1}{2\,\kappa} 
\,\pi^2-\frac{1}{2}\, 
\kappa\,\varphi'^2-2mM\cos{\sqrt{2}\varphi}+\frac{1}{\sqrt{2}}(-\pi 
A_1+\kappa\varphi'A_0)  \,   .  \ee 
On the equation of motion for the momentum $\pi$ 
\be \label {eq:jpi} 
\pi = \kappa (\dot{\varphi}-\frac{A_1}{\sqrt{2}}). 
\ee 
this Lagrangian takes a form 
\be \label {eq:flagr} 
\mathcal{L}^{Th}_{0}= \frac{1}{2}\,\kappa \,\partial_{\mu} \varphi \partial^{\mu}\varphi 
+\frac{\kappa}{\sqrt{2}}\,\epsilon^{\mu\nu} A_{\mu}\partial_{\nu}\varphi 
 -2mM \cos{(\sqrt{2}\,\varphi)}+ \frac{1}{4}\kappa A^2_1. 
\ee 
 
It is possible to add to the effective Lagrangian some local functional, 
depending on the fields $A_{+}$ and $A_{-}$. In order to obtain Lorentz 
invariant action of the effective bosonic theory, we will choose additional 
term in the form 
\bd 
\Delta\mathcal{L}^{Th}_{0}=-\frac{1}{4}\,\kappa  A^2_1. 
\ed 
Adding counterterm $\Delta\mathcal{L}^{Th}_{0}$ to the Lagrangian (\ref {eq:flagr}), and returning the term 
bilinear in $A_\mu$ we get 
\be \label{eq:eflag} 
\mathcal{L}^{Th}_{eff}= \frac{1}{2}\,\kappa \,\partial_{\mu} \varphi \partial^{\mu}\varphi 
 + \frac{\kappa}{\sqrt{2}}\,\epsilon^{\mu\nu} A_{\mu}\partial_{\nu}\varphi 
 -2mM \cos{(\sqrt{2}\,\varphi)}  + \frac{1}{8g}A^{\mu}A_{\mu}. 
\ee 
Now we can eliminate auxiliary field $A_{\mu}$ from Lagrangian (\ref {eq:eflag}), 
using its equation of motion 
\be \label {eq:ajna} 
A^\mu=- \frac{4g\kappa}{\sqrt{2}}\, 
\epsilon^{\mu\nu}\partial_\nu\varphi. 
\ee 
Substituting this equation back into Lagrangian (\ref {eq:eflag}), we get 
\be \label {eq:blagr} 
\mathcal{L}^{Th}_{eff}=\frac{1}{2}\,\kappa(1+2g\kappa)\,\partial_{\mu} \varphi 
\partial^{\mu}\varphi-2mM\cos{(\sqrt{2}\,\varphi)}. 
\ee 
This Lagrangian, after rescaling the scalar field $\varphi$, 
$\varphi\rightarrow  \frac{\sqrt{2}}{\beta} \varphi$, takes a form 
\be \label {eq:psg} 
\mathcal{L}^{Th}_{eff}=\frac{1}{2}\,\partial_{\mu} \varphi 
\partial^{\mu}\varphi-2mM\cos{(\beta\,\varphi)}, 
\ee 
where $\beta$ is defined by 
\be \label {eq:beta} 
\beta\equiv [\frac{1}{2}\kappa(1+2g\kappa)]^{-1/2}. 
\ee 
 
As it is usual, we will add constant term to the Lagrangian 
(\ref {eq:psg}), in order to have vanishing energy 
for the vacuum configuration $\varphi=0$ and obtain 
\be \label{eq:psgg} 
\mathcal{L}^{Th}_{eff}=\frac{1}{2}\,\partial_{\mu} \varphi 
\partial^{\mu}\varphi-2mM\,[\cos{(\beta\,\varphi)}-1]. 
\ee 
 
This is the Lagrangian of the sine--Gordon theory. Therefore, 
Thirring model is equivalent to the sine--Gordon theory, if there exists following relation 
between parameters 
\be \label {eq:ki} 
\frac{4\pi}{\beta^2\,\hbar}=1+\frac{g\hbar}{\pi}=1+2g\kappa  \,  , 
\ee 
which is consequence of (\ref {eq:beta}). Result given by Eq. (\ref {eq:ki}), obtained 
by canonical method, is in agreement with one in Ref. \cite{Kolman}. Coleman has obtained this 
result by direct computation of Green's functions in both Thirring and 
sine--Gordon  model using perturbative technique. 
The relation (\ref {eq:ki}) is the main result in this chapter. 
From its form, we can easy conclude that 
free Thirring model ($g=0$) is equivalent to sine--Gordon theory with 
\be \label {eq:free} 
\beta^2=\frac{4\pi}{\hbar}. 
\ee 
 
It is worth to emphasize that relation (\ref {eq:ki}) implies duality 
between Thirring and sine--Gordon model. Namely, from this relation it follows 
that the large values of Thirring coupling constant $g$ corresponds to the small 
value of sine--Gordon parameter $\beta$.
 
\subsection{One parameter class solutions of the massless Thirring model} 
We shall consider separately the massless Thirring model. This case is
specially interesting, because corresponding quantum theory is
non-uniquely defined. Namely, in the quantum action of the theory exists one
parameter, which does not appear in the classical one \cite{Hagen}. Since the
Thirring sine-Gordon relationship was already established, we will show the
existence of this parameter starting with the corresponding sine-Gordon model. 

Firstly, we split vector field from Lagrangian (\ref {eq:flagr}) to the
quantum and external part $A_\mu = a_\mu + A^{ex}_\mu$. 
Here the field $a_\mu$ plays the
role of our auxiliary field and $ A^{ex}_\mu$ is the Hagen's external source.
Then, omitting the
mass term and the local functional dependence on the vector fields, we obtain
from (\ref {eq:flagr})
\be
\mathcal{L}^{Th}(\varphi, a + A^{ex}) = \frac{1}{2}\,\kappa
\,\partial_{\mu} \varphi \partial^{\mu}\varphi + \frac{\kappa}{\sqrt{2}}\,\epsilon^{\mu \nu}
({a_{\mu} + A^{ex}_{\mu}) \partial_{\nu} \varphi.}
\ee

The invariance of the Thirring model under replacement
\begin{equation} \label{eq:repl}
j^\mu \to j^\mu_5 \,  , \qquad A^{ex}_\mu \to A^{ex}_{5 \mu} \,  , \qquad g \to -g,
\end{equation}
corresponds to the Lagrangian $\mathcal{L}^{Th}(\varphi_5, a_5 + A^{ex}_5)$.
Note that we introduce new auxiliary fields $\varphi_5$ and $a_5$, while the
external fields are related by the dual transformation $A^{ex}_{5 \mu}= \varepsilon_{\mu \nu}  A^{ex \nu}$.

The symmetry of the massless Thirring model given by  (\ref
{eq:repl}) allows us to introduce one-parameter Lagrangian 
\be \label{eq:opl}
\mathcal{L}(\xi, \eta) = \xi \mathcal{L}^{Th}(\varphi, a + A^{ex}) + \eta
\mathcal{L}^{Th}(\varphi_5, a_5 + A^{ex}_5) +\fr{1}{8 g}
(a_\mu a^\mu - a_{5 \mu} a_5^\mu)  \,  ,
\ee
with $\xi+\eta=1$. The parameters $\xi$ and $\eta$ obey this constraint because
the Lagrangian given by Eq. (\ref {eq:opl}) has to correspond
to the Lagrangian of the massless Thirring model. The terms quadratic in
auxiliary fields, in fact replaced the last term of the eq. (\ref {eq:eflag}).
The second part was adding with opposite sign in according to the symmetry
replacement $g \to -g$.

After elimination of the all auxiliary fields $a, a_5$ and then $\varphi,
\varphi_5$ we obtain the effective action
\be
W(A^{ex}) = - \fr{\hbar}{8} \int d^2 x A^{ex}_\mu D^{\mu \nu}_g A^{ex}_\nu \,  ,
\ee
where
\be
D^{\mu \nu}_g = \left( \varepsilon^{\mu\rho} \varepsilon^{\nu \sigma} {\xi \over 1+{g
\xi \hbar \over \pi}} + \eta^{\mu\rho} \eta^{\nu \sigma} {\eta \over 1-{g
\eta  \hbar \over \pi}} \right) \partial_\rho \partial_\sigma
\fr{1}{\partial^2}  \,   .
\ee
Up to the normalization factor it is just relation (3.13) from the Hagen's
paper Ref. \cite{Hagen}.

The expression for effective action is equivalent to the solution of the
functional integral, \cite{Sazdovic},
\be
<0 \mid 0>_{A,g}  = \int d {\bar \psi} d \psi e^{i \mathcal{L}_{Th}} = e^{i W(A^{ex})},
\ee
which corresponds to the Eq. (3.12) in the first Ref.
\cite{Hagen}. Using these expressions it is  easy to reproduce the other
Hagen's results.

\section{Bosonization of Fermionic Fields} 
 \cleq 
In this section we will apply canonical method of bosonization to the 
fermionic fields. Starting with PB of fermionic fields $\psi_{\pm}$ and 
corresponding momenta $\pi_{\pm}$ with currents $j_{\pm}$, fermionic fields 
will be expressed in terms of bosonic phase space coordinates $\varphi$ and 
$\pi$. After quantization, these classical fermionic fields become 
the 
operators. In order to show that these operators are really fermionic 
ones, we 
 investigate their anticommutation relations. From the bosonic form 
of the 
 operators $\Psi_{\pm}$, we easily obtain bosonic representation of 
scalar and 
 pseudoscalar density $\hat{\bar{\Psi}}\hat{\Psi}$ and 
$\hat{\bar{\Psi}}\gamma_5  \hat{\Psi}$, respectively. These results are 
consistent with ones obtained in Ref. \cite{Mandelstam}. 
 
\subsection{Construction of the fermionic field operators} 
 
Poisson brackets of the fermionic fields $\psi_{\pm}$ and its conjugate momenta 
$\pi_{\pm}$ with the currents $j_{\pm}$ have the form 
 \be \label {eq:pzpsi} 
\{j_{\pm}(\sigma), 
\psi_{\pm}(\bar{\sigma})\}=i\sqrt{2}\psi_{\pm}\delta(\sigma-\bar{\sigma}),\hspace*{.7cm} 
 \{j_{\pm}(\sigma),\psi_{\mp}(\bar{\sigma})\}=0, 
 \ee 
 \be \label {eq:pzpi} 
\{j_{\pm}(\sigma),\pi_{\pm}(\bar{\sigma})\}=-i\sqrt{2}\pi_{\pm}\delta(\sigma-\bar{\sigma}), 
\hspace*{.5cm} \{j_{\pm}(\sigma),\pi_{\mp}(\bar{\sigma})\}=0. 
\ee 
Because the right hand side is linear in the fields, the anomaly is absent and after quantization, 
algebra of the operators $\hat{\psi}_{\pm}$, $\hat{\pi}_{\pm}$ and $\hat{j}_{\pm}$ preserves the original form 
\be \label 
{eq:qalg}[\hat{j}_{\pm}(\sigma),\hat{\psi}_{\pm}(\bar{\sigma})]= 
-\hbar\sqrt{2}\hat{\psi}_{\pm}\delta(\sigma-\bar{\sigma}), 
\hspace*{.7cm} [\hat{j}_{\pm}(\sigma),\hat{\psi}_{\mp}(\bar{\sigma})]=0, 
\ee 
\be \label 
{eq:qalgg} [\hat{j}_{\pm}(\sigma),\hat{\pi}_{\pm}(\bar{\sigma})]= 
\hbar\sqrt{2}\hat{\pi}_{\pm}\delta(\sigma-\bar{\sigma}), 
\hspace*{1cm} [\hat{j}_{\pm}(\sigma),\hat{\pi}_{\mp}(\bar{\sigma})]=0. 
\ee 
 
Now, we will construct bosonic representation of the fermionic fields 
$\Psi_{\pm}$ and their conjugate momenta $\Pi_{\pm}$. We demand that 
Poisson brackets algebra of the fields $\Psi_{\pm}$, their conjugate momenta 
$\Pi_{\pm}$, and currents $J_{\pm}$, whose bosonic form is already known, is 
isomorphic to the algebra of the operators $\hat{\psi}_{\pm}$, 
$\hat{\pi}_{\pm}$ and $\hat{j}_{\pm}$, respectively. Therefore, we have 
\be \label {eq:bpz} 
\{J_{\pm}(\sigma), 
\Psi_{\pm}(\bar{\sigma})\}=i\sqrt{2}\Psi_{\pm}(\sigma)\delta(\sigma-\bar{\sigma}), 
\hspace*{.5cm}\{J_{\pm}(\sigma),\Psi_{\mp}(\bar{\sigma})\}=0, 
\ee 
\be \label {eq:bpzz} 
\{J_{\pm}(\sigma), 
\Pi_{\pm}(\bar{\sigma})\}=-i\sqrt{2}\Pi_{\pm}(\sigma)\delta(\sigma-\bar{\sigma}), 
\hspace*{.5cm}\{J_{\pm}(\sigma),\Pi_{\mp}(\bar{\sigma})\}=0. 
\ee 
In order to solve these equations in terms of $\Psi_{\pm}$ and $\Pi_{\pm}$, 
let us introduce the variables $I_{\pm}$ as follows 
\be \label {eq:i} 
I_{\pm}(\sigma)\equiv\int_{-\infty}^{\sigma}\!d\sigma_1\, 
J_{\pm}(\sigma_1).\ee 
With the help of bosonic representation of currents (\ref {eq:sjpm}), 
we obtain the $I_{\pm}$ dependence of basic bosonic variables 
\be \label {eq:ifi} 
I_{\pm}(\sigma)=\pm \int_{-\infty}^{\sigma}\!d\sigma_1\, 
\pi(\sigma_1)+\kappa\varphi(\sigma). 
\ee 
Using Poisson brackets current algebra, given by Eq.(\ref {eq:alj}), we 
find PB of the variables $I_{\pm}$ with the currents $J_{\pm}$ 
\be \label {eq:jial} 
\{J_{\pm}(\sigma),I_{\pm}(\bar{\sigma})\}=\mp 
2\kappa\,\delta(\sigma-\bar{\sigma}), 
\ee 
\be \label {jiall} 
\{J_{\pm},I_{\mp}\}=0. 
\ee 
Let us suppose that fields $\Psi_{\pm}$ and their conjugate momenta depend 
only on variables $I_{\pm}$. Under that assumption, from the algebra given by 
Eqs.(\ref {eq:bpz}) and (\ref {eq:bpzz}), we get the following equations 
\be 
\{J_{\pm}(\sigma),I_{\pm}(\bar{\sigma})\}\frac{\partial\Psi_{\pm}}{\partial 
I_{\pm}}=i\sqrt{2}\Psi_{\pm}(\sigma)\delta(\sigma-\bar{\sigma}), 
\ee 
\be 
\{J_{\pm}(\sigma),I_{\pm}(\bar{\sigma})\}\frac{\partial\Pi_{\pm}}{\partial 
I_{\pm}}=-i\sqrt{2}\Pi_{\pm}(\sigma)\delta(\sigma-\bar{\sigma}). 
\ee 
With the help of (\ref {eq:jial}), we obtain bosonic representation of fermionic field 
$\Psi_{\pm}$, and their conjugate momenta $\Pi_{\pm}$ 
\be 
\label {eq:aabpsi} 
\Psi_{\pm}= C_{\pm}\exp{(\mp 
\frac{i}{\kappa\sqrt{2}}\,I_{\pm})},\ee \be \label {eq:abpi} 
\Pi_{\pm}= D_{\pm}\exp{(\pm \frac{i}{\kappa\sqrt{2}}\,I_{\pm})}, 
\ee 
where $C_{\pm}$ and $D_{\pm}$ are the constants which will be determined using regularization 
procedure. These constants are not independent. 
The relation $D_{\pm}=i\,C^*_{\pm}$ follows from classical fermionic theory constraints 
$\pi_{\pm}=i \psi^*_{\pm}$ and its bosonic analogue 
$\Pi_{\pm}=i \Psi^*_{\pm}$. 
 
After quantization the classical fields $\Psi_{\pm}$ and their conjugate momenta 
$\Pi_{\pm}$ become operators $\hat{\Psi}_{\pm}$ and $\hat{\Pi}_{\pm}$. So, 
normal ordering prescription has to be applied to the 
operators product in the right hand side in Eqs.(\ref {eq:aabpsi}) and 
(\ref {eq:abpi}) 
\be \label 
{eq:qbpsi}\hat{\Psi}_{\pm}= C_{\pm}:\exp{(\mp 
\frac{i}{\kappa\sqrt{2}}\,\hat{I}_{\pm})}:\, , 
\ee 
\be \label 
{eq:qbpi}\hat{\Pi}_{\pm}= i\,C^*_{\pm}:\exp{(\pm 
\frac{i}{\kappa\sqrt{2}}\,\hat{I}_{\pm})}:\, . 
\ee 
This means that after expansion of the exponent in the right hand side, 
annihilation operators are placed right to the creation ones (Appendix A). 
 
As direct computation shows (Appendix B), products of the operators 
$\hat{\Psi}^*_{\pm}\hat{\Psi}_{\pm}$ at the same point of space are singular. 
In order to regularize these products, let us introduce the operators 
$\hat{\tilde{J_{\pm}}}$, which are the product of field operators at the 
different points of space 
\be \label {eq:nqbpsi} 
\hat{\tilde{J_{\pm}}}(\sigma_1,\sigma_2) 
\equiv\sqrt{2}\hat{\Psi}^*_{\pm}(\sigma_1) 
\hat{\Psi}_{\pm}(\sigma_2). 
\ee 
After some calculations (Appendix B), we obtain 
\be \label {eq:regj} 
\hat{\tilde{J}}_{\pm}(\sigma,\sigma+\eta)|\,_{\eta\rightarrow 0}\equiv 
\sqrt{2}\hat{\Psi}^*_{\pm}(\sigma)\hat{\Psi}_{\pm}(\sigma+\eta)|\,_{\eta\rightarrow0} 
=\frac{F_{\pm}}{\eta\pm i \varepsilon}+Z_{\pm} 
\hat{J}_{\pm}(\sigma),\,\,(\varepsilon>0), 
\ee 
where $F_{\pm}$ and $Z_{\pm}$ are given by following expressions 
($\Lambda$--cutoff parameter) 
\be 
F_{\pm} = \pm i\Lambda\sqrt{2}|C_{\pm}|^2, 
\hspace*{1cm} Z_{\pm}=\frac{\Lambda}{\kappa}|C_{\pm}|^2. 
\ee 
Because $\hat{\Psi}_{\pm}$ is a representation of the $\hat{\psi}_{\pm}$, we expect that 
bilinear combination in (\ref {eq:regj}) produce $\hat{J}_{\pm}$, since it is the same combination 
as in (\ref {eq:j}). So, the natural choice for the constants is 
\bd 
Z_{\pm}=1. 
\ed 
From the last relation, we get the values of the constants $F_{\pm}$ and $C_{\pm}$ 
\be 
C_{\pm}=\sqrt{\frac{\kappa}{\Lambda}}\,\,,\hspace*{1cm} 
F_{\pm}=\pm i \kappa \sqrt{2}. 
\ee 
With that values of the constants $F_{\pm}$ and $C_{\pm}$, the operators 
$\hat{\Psi}_{\pm}$, $\hat{\Pi}_{\pm}$ and $\hat{\tilde{J}}_{\pm}$ take a form 
\be \label {eq:kpsi} 
\hat{\Psi}_{\pm}= \sqrt{\frac{\kappa}{\Lambda}}:\exp{(\mp 
\frac{i}{\kappa\sqrt{2}}\hat{I}_{\pm})}:\,\, , 
\ee 
\be \label 
{eq:kpi}\hat{\Pi}_{\pm}= i \sqrt{\frac{\kappa}{\Lambda}}:\exp{(\pm 
\frac{i}{\kappa\sqrt{2}}\hat{I}_{\pm})}:\,\, , 
\ee 
\be \label 
{eq:kj} \hat{\tilde{J}}_{\pm}(\sigma, 
\sigma+\eta)|_{\eta\rightarrow 0}=\pm 
\frac{i\kappa\sqrt{2}}{\eta\pm i 
\varepsilon}+\hat{J}_{\pm}(\sigma). 
\ee 
 
In order to compare these results with ones from the Ref. \cite{Mandelstam}, 
we will derive Lagrangian form of the operators $\hat{\Psi}_{\pm}$. Note that 
up to this point in Sec. 3. we did not specify Hamiltonian of the theory. So, 
our canonical expression is valid for any theory satisfying PB algebra (\ref 
{eq:pzpsi}, \ref {eq:pzpi}). Passing to the Lagrangian formulation we must be 
more specific and we chose the example 
 of the Thirring model, which we 
considered in the previous section. We will express the momentum $\pi$ in 
terms of corresponding velocity 
starting from equation of motion for the 
momentum, Eq. (\ref 
 {eq:jpi}), and using equation of motion for auxiliary 
field $A_{\mu}$ 
 \be \label {eq:timpuls} 
\pi =  \frac{2}{\beta^2}\dot{\varphi}. 
\ee 
Substituting the last equation in Eq. (\ref {eq:aabpsi}), after rescaling 
$\varphi \rightarrow  \frac{\sqrt{2}}{\beta}\varphi$, we obtain Lagrangian 
form of the 
 fields $\Psi_{\pm}$ 
\be 
\Psi_{\pm}(\sigma)=C_{\pm}\exp{\{-\frac{i}{\kappa\beta}\int_{-\infty}^{\sigma}\!d\bar{\sigma} 
\dot{\varphi}(\bar{\sigma})\mp \frac{i\beta}{2}\varphi(\sigma)\}}. 
\ee 
After quantization, from the last relation, we get 
\be \label {eq:lkvpsi} 
\hat{\Psi}_{\pm}(\sigma)=C_{\pm}\,:\exp{\{-\frac{i}{\kappa\beta} 
\int_{-\infty}^{\sigma}\!d\bar{\sigma} 
\hat{\dot{\varphi}}(\bar{\sigma})\mp 
\frac{i\beta}{2}\hat{\varphi}(\sigma)\}}:\,\,. 
\ee 
 
This form of the field operators 
is in agreement with one obtained in Ref. \cite{Mandelstam} from the requirement that operators 
$\hat{\Psi}_{\pm}$ have to be annihilation operators for solitons in 
sine--Gordon theory, as well as to anticommute with itself. Result given by 
Eq. (\ref {eq:kpsi}) is more general, because it is in Hamiltonian form, so 
it can be applied to the other two--dimensional models. Additionally, this 
result is obtained from the less number of assumptions. Namely, we obtained 
this result demanding that fields $\Psi_{\pm}$ and currents $J_{\pm}$ have to 
obey Poisson brackets algebra which is isomorphic to algebra of the operators 
$\hat{j}_{\pm}$ and $\hat{\psi}_{\pm}$. 
 
\subsection{Anticommutation relations for the operators ${\hat \Psi}_\pm$ and ${\hat \Pi}_\pm$} 
 
In this subsection we will show that operators $\hat{\Psi}_{\pm}$ and $\hat{\Pi}_{\pm}$, 
given by (\ref {eq:kpsi}) i  (\ref {eq:kpi}), obey canonical anticommutation relations. 
 
In order to justify interpretation of the operators $\hat\Psi_{\pm}$ and $\hat\Pi_{\pm}$ 
as the fermionic operators, we should show that they obey canonical anticommutation relations. 
\ Firstly, we will find anticommutation relations for the operators 
$\hat{\Psi}_{\pm}$. Using Eqs. (\ref {eq:kpsi}) and (\ref {eq:no}), we find 
\be 
\hat{\Psi}_{\pm}(\sigma)\hat{\Psi}_{\pm}(\bar{\sigma})= 
\frac{\kappa}{\Lambda}\exp{\{-\frac{1}{2\kappa^2}\, 
[\hat{I}^{(\pm)}_{\pm}}(\sigma), 
\hat{I}^{(\mp)}_{\pm}(\bar{\sigma})]\}:\exp{\{\mp\frac{1}{2\kappa^2}[\hat{I}_{\pm}(\sigma)+ 
\hat{I}_{\pm}(\bar{\sigma})]\}}:\,\,.\ee 
 
With the help of Eq. (\ref {eq:expmX}), in the limit $\varepsilon\rightarrow0$, we obtain that 
anticommutator for the fields $\Psi_{\pm}$ vanish 
\be \label {eq:apsi} 
[\hat{\Psi}_{\pm}(\sigma),\hat{\Psi}_{\pm}(\bar{\sigma})]_{+}=0.\ee 
The calculation, which is very similar to the previous one, shows that the 
anticommutator for the momenta also vanish 
\be \label {eq:api} 
[\hat{\Pi}_{\pm}(\sigma),\hat{\Pi}_{\pm}(\bar{\sigma})]_{+}=0.\ee 
 
Now we will find anticommutator for the fields $\Psi_{\pm}$ with their 
conjugate momenta $\Pi_{\pm}$. Using Eqs. (\ref {eq:kpsi}), (\ref {eq:kpi}), 
and (\ref {eq:no}), we obtain 
 \bd 
\hat{\Psi}_{\pm}(\sigma)\hat{\Pi}_{\pm}(\bar{\sigma})= 
i\,\frac{\kappa}{\Lambda}\,\exp{\{\frac{1}{2\kappa^2}\, 
[\hat{I}^{(\pm)}_{\pm}}(\sigma), 
\hat{I}^{(\mp)}_{\pm}(\bar{\sigma})]\} 
:\exp{\{\mp\frac{i}{\kappa\sqrt{2}}[\hat{I}_{\pm}(\sigma)-\hat{I}_{\pm}(\bar{\sigma})]\}}: \,  , 
\ed 
and with the help of Eq. (\ref {eq:expX}), we get 
\bd 
\hat{\Psi}_{\pm}(\sigma)\hat{\Pi}_{\pm}(\bar{\sigma})= 
i\hbar\,\delta^{(\pm)}(\sigma-\bar{\sigma}) 
:\exp{\{\mp\frac{i}{\kappa\sqrt{2}}[\hat{I}_{\pm}(\sigma)-\hat{I}_{\pm} 
(\bar{\sigma})]\}}:\, , 
\ed 
which produces canonical anticommutation relations for the operators 
$\hat{\Psi}_{\pm}$ and $\hat{\Pi}_{\pm}$ 
 \be 
\label {eq:apipsi} 
[\hat{\Psi}_{\pm}(\sigma),\hat{\Pi}_{\pm}(\bar{\sigma})]_{+}= 
i\hbar \delta(\sigma-\bar{\sigma}) 
:\exp{\{\mp\frac{i}{\kappa\sqrt{2}}[\hat{I}_{\pm}(\sigma)-\hat{I}_{\pm}(\bar{\sigma})]\}}:\,= 
i\hbar\delta(\sigma-\bar{\sigma}). 
\ee 
 
\subsection{Bosonization of the scalar densities} 
 
Products of the fermionic field operators $\hat{\bar{\Psi}}\hat{\Psi}$ and 
$\hat{\bar{\Psi}}\gamma^5\hat{\Psi}$ we will express in terms of bosonic 
variables using Eq. (\ref {eq:kpsi}). Note that the relation $[{\hat I}_+ , 
{\hat I}_-]=0$ 
 simplifies the calculations. With the help of Eq. (\ref {eq:no}), 
we 
 obtain 
 \be 
\hat{\Psi}^*_\pm \hat{\Psi}_\mp =\frac{\kappa}{\Lambda} 
:\exp{[\frac{\pm i}{\kappa\sqrt{2}}(\hat{I}_{+}+\hat{I}_{-})]}:\,\,. 
\ee 
Substituting Eq. (\ref {eq:ifi}) in the last equation, we find 
\be 
\hat{\Psi}^*_\pm \hat{\Psi}_\mp =\frac{\kappa}{\Lambda}\,:\exp{(\pm i\sqrt{2}\varphi)}:\,\,. 
\ee 
So, the bosonic representations for scalar density 
$\hat{\bar{\Psi}}\hat{\Psi}$ and pseudoscalar density $\hat{\bar{\Psi}}\gamma^5\hat{\Psi}$ are 
\be 
\hat{\bar{\Psi}}\hat{\Psi}=\hat{\Psi}_{-}^*\hat{\Psi}_{+}+ \hat{\Psi}_{+}^*\hat{\Psi}_{-} 
=\frac{\hbar}{\pi\Lambda}:\cos{(\sqrt{2}\varphi)}:\,\,, 
\ee 
\be 
\hat{\bar{\Psi}}\gamma^5\hat{\Psi}=-\hat{\Psi}_{-}^*\hat{\Psi}_{+}+\hat{\Psi}_{+}^*\hat{\Psi}_{-} 
=\frac{i\hbar}{\pi\Lambda}:\sin{(\sqrt{2}\varphi)}:\,\,. 
\ee 
These results are consistent with ones obtained by direct 
applying of the method to the scalar densities. 
 
\section{Conclusion} 
 
In this paper we presented a complete and independent derivation of the Thirring sine-Gordon 
relationship, using the Hamiltonian methods. We also obtained 
Hamiltonian and Lagrangian  representation for the Mandelstam's fermionic operators . 
 
We started with canonical analysis of the theory where 
fermions are coupled to the auxiliary external gauge field. The massive Thirring model 
can be easily obtained from this Lagrangian by adding square of the auxiliary field and 
eliminating it on its equation of motion. We found that there exists FCC $j_{\pm}$ in our theory, 
whose  PB are equal to zero. In the quantum theory, the central term appears in the commutation relations 
of the operators ${\hat j}_{\pm}$. This changes the nature of constraints because they become SCC. 
 
We define the new effective theory, postulating the PB of the constraints 
and Hamiltonian density, following the method developed in \cite{Sazdovic}. 
We require that the classical PB algebra of the bosonic theory is isomorphic to the quantum 
commutator algebra of the fermionic theory. Then we found the 
representation for the currents and Hamiltonian density in terms of 
phase space coordinates. Finally, we derived effective action using 
general canonical formalism and obtained the equivalent bosonizated model. 
Together with auxiliary field term this is just the sine-Gordon action, up to some 
parameters identifications which agrees with Refs. \cite{Kolman,Mandelstam}.
For the massless Thirring model it is shown that
its quantum effective action has one-parameter which
does not exist in the classical one. We determined the quantum action using
formal invariance of the massless Thirring model under replacements given by
Eq. (\ref {eq:repl}) and already established Thirring sine-Gordon relationship.

The algebra of the currents $J_{\pm}$ is the  basic
PB algebra.  Knowing the representation of the currents $J_{\pm}$ in terms of
$\varphi$ and $\pi$  we can find the representation for all other quantities
from their  PB algebra with the currents. In Sec. 2. we found the bosonization
rules  for the chiral densities. The main result of Sec. 3. is the bosonic 
representation for   the fermions, which has been obtained in the same way.   
Beside usual bosonization rules and usual Mandelstam's fermionic
representations,  we also got the Hamiltonian ones, expressing the currents
$J_{\pm}$  in terms of both coordinate $\varphi$ and momentum $\pi$.  These
rules are more general, because they are valid for arbitrary Hamiltonian  and
they are  consequence of the commutation relations. After momenta  elimination
on their equations of motion, we came back to the conventional bosonization 
rules and to the conventional Mandelstam's fermionic representations.   
The Schwinger term, and consequently the sine-Gordon action 
have the correct dependence on Planck's constant $\hbar $, because $\k$ 
is proportional to $\hbar$. The fact that $\hbar$ arises in the classical 
effective theory and  in the coupling constant relation, shows the quantum 
origin of the established equivalence. 
\appendix 
 
\section{Normal Ordering and Central Term} 
 \cleq 
In this appendix we will derive commutation relations of the current 
operators\\ $\hat{j}_{\pm}\equiv-i\sqrt{2}:\hat{\pi}_{\pm}\hat{\psi}_{\pm}:$ 
\be \label {eq:jalg} 
[\hat{j}_{\pm}(\sigma),\hat{j}_{\pm}(\bar{\sigma})]=\pm 2 i\hbar\kappa 
\delta'(\sigma-\bar{\sigma}),\hspace*{1cm}[\hat{j}_{\pm}(\sigma), 
\hat{j}_{\mp}(\bar{\sigma})]=0,\ee 
where $\kappa\equiv\frac{\hbar}{2\pi}$. 
 
The current operators $\hat{j}_{\pm}$ we define using 
normal ordering prescription. In order to decompose these operators in positive 
and negative frequencies in position space, let us introduce two parts of the 
delta function 
\be \label {eq:pndelta} 
\delta^{(\pm)}(\sigma)=\int_{-\infty}^{\infty}\!\frac{dk}{2\pi}\, 
\theta(\mp k)\,e^{ik(\sigma \mp i\varepsilon)}\,=\,\frac{\mp 
i}{2\pi(\sigma \mp i\varepsilon )},\;(\varepsilon > 0), 
\ee 
where $\theta$ is unit step function. They obviously obey the relation $\delta(\sigma)=\delta^{(+)}(\sigma)+ 
\delta^{(-)}(\sigma)$ and have the following properties 
\be 
\delta^{(+)}(\sigma)=\delta^{(-)}(-\sigma), 
\ee 
\be \label {eq:dva} 
[\delta^{(+)}(\sigma)]^2 - 
[\delta^{(-)}(\sigma)]^2=\frac{i}{2\pi}\,\delta'(\sigma). 
\ee 
Any operator $\hat{O}$  can also be decomposed in two parts 
\be 
\hat{O}^{(\pm)} 
(\tau,\sigma)=\,\int_{-\infty}^{\infty}\!d\bar{\sigma}\,\delta^{(\pm)}(\sigma-\bar{\sigma})  \, \hat{O}(\tau,\bar{\sigma}), 
\ee 
so that $\hat{O}=\hat{O}^{(+)}+\hat{O}^{(-)}$.  We promote the operators 
$\hat{\pi}_{+}^{(-)}$ and  $\hat{\psi}_{+}^{(-)}$ as annihilation ones, and 
the operators 
 $\hat{\pi}_{+}^{(+)}$ and $\hat{\psi}_{+}^{(+)}$ as creation 
ones 
 \be 
\hat{\pi}_{+}^{(-)}|\,0\rangle = \hat{\psi}_{+}^{(-)}|\,0 \rangle=0 \,  , 
\qquad 
 0|\,\hat{\pi}_{+}^{(+)}=\langle0|\,\hat{\psi}_{+}^{(+)}=0   \,  . 
\ee 
In order to preserve the simmetry under parity transformations, we define 
creation and annihilation operators for $\hat{\pi}_{-}$ i $\hat{\psi}_{-}$ in 
an oposite way [operators with index $(-)$ are creation and ones with index 
$(+)$ are annihilation]. Normal order for product of the operators means 
that  creation operators are placed to the left from the annihilation ones. 
 
From the basic commutation relations 
\be 
[\hat{\psi}_{\pm}(\sigma),\hat{\pi}_{\pm}(\bar{\sigma})]=\,i\hbar\, 
\delta(\sigma-\bar{\sigma}), 
\ee 
we have 
\be \label {eq:pmalg} 
[\hat{\psi}_{+}^{(\pm)}(\sigma),\hat{\pi}_{+}^{(\mp)}(\bar{\sigma})]= 
i\hbar\,\delta^{(\pm)} (\sigma-\bar{\sigma}),\ee \be 
[\hat{\psi}_{-}^{(\pm)}(\sigma),\hat{\pi}_{-}^{(\mp)}(\bar{\sigma})]=i\hbar\,\delta^{(\pm)} 
(\sigma-\bar{\sigma}), 
\ee 
and other commutation relations are trivial. Since 
the Poisson brackets for the currents $j_{\pm}$ vanish, only possible 
difference between classical and quantum algebra is appearance of central term 
at the quantum level. Because of that, we find the form of the current algebra taking 
vacuum expectation value of the commutators 
\be 
[\hat{j}_{\pm}(\sigma),\hat{j}_{\pm}(\bar{\sigma})]= 
\Delta_{\pm}(\sigma,\bar{\sigma})- \Delta_{\pm}(\bar{\sigma},\sigma),\ee \ 
where $\Delta_{\pm}(\sigma,\bar{\sigma})\equiv 
\langle0|\,\hat{j}_{\pm}(\sigma) \hat{j}_{\pm}(\bar{\sigma})|\,0\rangle$. 
 
Using the fact that the operators $\hat{j}_\pm$ are normal ordered, the only nontrivial 
contributions have the form 
\be \label {eq:ap} 
 \Delta_{\pm}(\sigma,\bar{\sigma})=-2 
\langle0|\,\hat{\pi}_{\pm}^{(\mp)}(\sigma) 
  \hat{\psi}_{\pm}^{(\mp)}(\sigma) 
\hat{\pi}_{\pm}^{(\pm)}(\bar{\sigma})\hat{\psi}_{\pm}^{(\pm)}(\bar{\sigma})|\,0\rangle= 
-2\hbar^2\,[\delta^{(\mp)}(\sigma-\bar{\sigma})]^2. 
\ee 
With the help of the $\delta^{(\pm)}$ functions properties (\ref {eq:dva}) 
we obtain Eq. (\ref {eq:jalg}). Commutator for the 
currents with different lower indices do not have central term, as well as the 
commutators of the currents with the operators $\hat{t}_{\pm}$ and 
$\hat{\rho}_{\pm}$. 
 
\section{Regularization of the Field Products} 
 \cleq 
In this Appendix, we will show, using regularization procedure, that following 
relations hold 
\be \label {eq:p} 
\sqrt{2}\hat{\Psi}^*_{\pm}(\sigma)\hat{\Psi}_{\pm}(\sigma+\eta)|_{\eta\rightarrow0} 
\equiv\hat{\tilde{J}}_{\pm}(\sigma,\sigma+\eta)|_{\eta\rightarrow 
0}=\frac{F_{\pm}}{\eta\pm i \varepsilon}+Z_{\pm} 
\hat{J}_{\pm}\,,         \qquad  (\varepsilon>0) 
\ee 
where $F_{\pm}$ i $Z_{\pm}$ are given by ($\Lambda$--cutoff parametar) 
\be \label {eq:const} 
F_{\pm} = \pm i\Lambda\sqrt{2}\,|C_{\pm}|^2 \, , 
\hspace*{1cm} Z_{\pm}=\frac{\Lambda}{\kappa}\,|C_{\pm}|^2 \, . 
\ee 
Starting with the definition of the operator $\hat{\tilde{J}}_{\pm}$ and 
using formula (Eq. (3.5) in Ref. \cite{Mandelstam}) 
\be \label {eq:no} 
:e^{\hat{A}}:\,:e^{\hat{B}}=e^{[\hat{A}^{(+)}\hat{B}^{(-)}]}:e^{\hat{A}+\hat{B}}:\,, 
\ee 
($[A^{(+)},B^{(-)}]$ -- c -- number), we get 
\be \label {eq:tj} 
\hat{\tilde{J}}_{\pm}(\sigma,\bar{\sigma})=\sqrt{2}|C_{\pm}|^2 
\exp{\{\frac{1}{2\kappa^2}\,X_{\pm}(\sigma,\bar{\sigma})\}}:\exp{\{\pm\frac{i}{\kappa\sqrt{2}} 
[\hat{I}_{\pm}(\sigma)-\hat{I}_{\pm}(\bar{\sigma})]\}}:\,, 
\ee 
where 
\be \label {eq:ex} 
X_{\pm}(\sigma,\bar{\sigma})\equiv [\hat{I}_{\pm}^{(\pm)}(\sigma),\hat{I}_{\pm}^{(\mp)}(\bar{\sigma})]. 
\ee 
We compute the commutators $X_{\pm}$ using the algebra of the operators $\hat{I}_{\pm}$ 
\be 
[\hat{I}_{\pm}(\sigma), \hat{I}_{\pm}(\bar{\sigma})]=\int_{-\infty}^{\sigma}\!d\sigma_1 
\int_{-\infty}^{\bar{\sigma}}\!d\sigma_2 
[\hat{J}_{\pm}(\sigma_1),\hat{J}_{\pm}(\sigma_2)]. 
\ee 
The last relation can be rewritten in a form 
\be \label {eq:pom} 
[\hat{I}_{\pm}(\sigma),\hat{I}_{\pm}(\bar{\sigma})]=\pm i\hbar\kappa 
\int_{-\infty}^{\sigma}d\sigma_1\!\int_{-\infty}^{\bar{\sigma}}\!d\sigma_2\, 
[\partial_{\sigma_1} \delta(\sigma_1-\sigma_2)-\partial_{\sigma_2} 
\delta(\sigma_1-\sigma_2)], 
\ee 
which is obtained from algebra of the currents $\hat{J}_{\pm}$. Performing 
integration we get 
\be 
\label {eq:ialg} 
[\hat{I}_{\pm}(\sigma),\hat{I}_{\pm}(\bar{\sigma})]=\mp 
i\hbar\kappa \varepsilon(\sigma-\bar{\sigma}), 
\ee 
where $\varepsilon(\sigma)=\theta(\sigma)-\theta(-\sigma)$ is a sign function.
From the expression 
\be 
[\hat{I}_{\pm}^{(\pm)},\hat{I}_{\pm}]=[\hat{I}_{\pm}^{(\pm)},\hat{I}_{\pm}^{(\mp)}], 
\ee 
follows that 
\be \label {eq:iali} 
X_{\pm}(\sigma,\bar{\sigma})\equiv 
[\hat{I}_{\pm}^{(\pm)}(\sigma),\hat{I}_{\pm}^{(\mp)}(\bar{\sigma})]= 
\mp 
i\hbar\kappa\int_{-\infty}^{\infty}d\sigma_1\,\delta^{(\pm)}(\sigma-\sigma_1)\, 
\varepsilon(\sigma_1-\bar{\sigma}). 
\ee 
Using Eq. (\ref {eq:pndelta}), the last relation obtains the form 
\be \label {eq:alii} 
X_{\pm}= \mp i\hbar \kappa\int_{-\infty}^{\infty}\! d\sigma_1\,  \frac{\mp 
i}{2\pi(\sigma - \sigma_1\mp i \varepsilon)} \, 
\varepsilon(\sigma_1-\bar{\sigma})   \,    . 
\ee 
After regularization  ($\Lambda$--cutoff parameter) integral on the right hand side 
takes a form 
\be 
X_{\pm}=-\kappa^2\left\{ 
\int_{\bar{\sigma}}^{\Lambda}\!d\sigma_1\,\frac{1}{\sigma-\sigma_1\mp i 
\varepsilon}- \int_{-\Lambda}^{\bar{\sigma}}\!d\sigma_1 \frac{1}{\sigma-\sigma_1\mp i 
\varepsilon}\right\}  \,   . 
\ee 
Computing the integral and taking $\Lambda \gg \sigma$, 
we get 
\be 
X_{\pm}=\kappa^2\left\{\ln{[\frac{\Lambda^2}{(\bar{\sigma}-\sigma\pm i 
\varepsilon)^2}]}\pm i \pi\right\}. 
\ee 
The last equation implies 
\be 
\label {eq:expX} 
e^{\frac{1}{2\kappa^2}X_{\pm}(\sigma,\, \bar{\sigma})}=\,\frac{\mp i 
\Lambda}{\sigma-\bar{\sigma} \mp i 
\varepsilon} = 2\pi\Lambda\delta^{(\pm)} 
(\sigma-\bar{\sigma}), 
\ee 
\be \label {eq:expmX} 
e^{-\frac{1}{2\kappa^2}X_{\pm}(\sigma,\, \bar{\sigma})}=\pm i\, 
\frac{\sigma-\bar{\sigma}\mp i \varepsilon}{\Lambda}  \,  . 
\ee 
Substituting Eq. (\ref {eq:expX}) in Eq. (\ref {eq:tj}), we get the 
regularized form of the operator $\hat{\tilde{J}}_{\pm}$\\ 
\be \label 
{eq:rtj}\hat{\tilde{J}}_{\pm}(\sigma,\bar{\sigma})=\sqrt{2}|C_{\pm}|^2 
\frac{\mp i \Lambda}{\sigma-\bar{\sigma}\mp i \varepsilon 
}:\exp{\{\pm\frac{i}{\kappa\sqrt{2}} 
[\hat{I}_{\pm}(\sigma)-\hat{I}_{\pm}(\bar{\sigma})]\}}:\, 
,\,\,\,(\varepsilon>0). 
\ee 
For the infinitesimal $\eta ={\bar \sigma}- \sigma$ we get 
\be \label 
{eq:rrtj} 
\hat{\tilde{J}}_{\pm}(\sigma,\sigma+\eta)|_{\eta\rightarrow 0}= 
\sqrt{2}\,|C_{\pm}|^2\,\frac{\pm i\,\Lambda}{\eta\pm i \varepsilon 
}\,+\,\frac{|C_{\pm}|^2\,\Lambda}{\kappa}\,\hat{J}_{\pm},\,\,\,\,(\varepsilon>0)\,, 
\ee 
which is exactly the relation (\ref {eq:p}), and $F_{\pm}$ and $C_{\pm}$ 
are given by Eq. (\ref {eq:const}). 
\section*{Acknowledgments} 
Work supported in part by the Serbian Ministry of Science, Technology 
and Development, under contract No. 1486. V.J. was partly supported by Swiss 
National Foundation under grant No. 620-62868.00.



\end{document}